\documentclass[aps,prd,twocolumn,nofootinbib,superscriptaddress]{revtex4-2}
\usepackage[T1]{fontenc}
\usepackage{amsmath,amssymb,amsthm,mathtools,bm}
\usepackage{graphicx}
\usepackage{xcolor}
\usepackage[hidelinks]{hyperref}
\usepackage{lmodern}
\newcommand{\dd}{\mathrm{d}}

\newcommand{\Sch}{\mathrm{Sch}}

\newcommand{\Ei}{\operatorname{Ei}}
\begin{document}
\title{Extended spherically symmetric solutions in revised Deser--Woodard nonlocal gravity}
\author{Haida Li}
\affiliation{School of Physics and Optoelectronics, South China University of Technology, Guangzhou 510641, China}
\author{Xiangdong Zhang}
\email{scxdzhang@scut.edu.cn}
\affiliation{School of Physics and Optoelectronics, South China University of Technology, Guangzhou 510641, China}
\begin{abstract}
In this work, we extend the static spherically symmetric black hole solutions of revised Deser--Woodard (D-W) nonlocal gravity. Due to the linearity of the field equations, we show that the first-order expansion around the Schwarzschild solution is linear in both the temporal metric component and the reciprocal of the radial metric component. Therefore, a mode-by-mode superposition can be properly defined for different types of correction terms. We then further study two additional asymptotically flat extensions, a logarithmically dressed correction and an exponentially suppressed correction. The logarithmically dressed correction gives a slowly decaying deviation from Schwarzschild as the radius becomes large, making it a more extended nonlocal effect, whereas the exponentially suppressed correction is localized near the horizon and shifts the horizon inward for positive correction amplitude. These correction terms are physically motivated by common mechanisms in modified gravity: Logarithmic terms are possibly related to effects of quantum corrections, and exponential terms can arise from finite-range or screened gravitational effects.
\end{abstract}
\maketitle
\section{Introduction}
Although general relativity explains gravity through the curvature of spacetime and has been extensively validated through various experiments, at the moment, it still lacks a fully consistent quantum formulation aligned with the other fundamental forces, and its theoretical limitations become especially important in extreme gravity regimes, such as the vicinity of black holes (BHs) \cite{Ashtekar:2004eh,Will:2014kxa,LIGOScientific:2016aoc,EventHorizonTelescope:2019dse,Ishak:2018his,Barack:2018yly,DAgostino:2023cgx,Li:2024afr,Li:2025mcp,Li:2025zdt}. On cosmological scales, the $\Lambda$CDM model contains dark energy and dark matter, while the cosmological constant is tied to fine-tuning issues when interpreted as vacuum energy predicted by quantum field theory. These tensions motivate alternative scenarios aimed at the interplay between gravity and quantum mechanics \cite{SupernovaCosmologyProject:1998vns,SupernovaSearchTeam:1998fmf,Peebles:2002gy,Frieman:2008sn,DAgostino:2019wko,Weinberg:1988cp,Padmanabhan:2002ji,DAgostino:2022fcx,Planck:2018vyg}.

One promising approach is modifying the theory of gravity by introducing nonlocal terms, which account for the influence of the entire spacetime. A well-known example of a nonlocal gravity scenario is the Deser--Woodard (D-W) model, proposed to reproduce the late-time history of the Universe without the fine-tuning issues of the cosmological constant \cite{Deser:2007jk}. The original model was later found to fail in predicting small-scale observations due to the lack of a suitable screening mechanism in the solar system, and the refined, revised D-W model was proposed to address this limitation \cite{Deser:2019lmm}. This improved version has been applied to cosmology and BH quasinormal modes \cite{Nojiri:2007uq,Calcagni:2007ru,Biswas:2011ar,Park:2012cp,Maggiore:2014sia,Dirian:2014ara}.

Static and spherically symmetric solutions can provide a useful ground for testing the nonlocal effects of revised D-W gravity. The black-hole solution obtained in \cite{DAgostino:2025wgl} gives a concrete starting point: by reformulating the vacuum Einstein field equations in a suitable tetrad frame, it reduces the complexity of these equations and produces an analytic solution with a \(1/r^n\) correction to the temporal metric component, a corresponding correction to the radial component up to first order in the coupling parameter, and a reconstructed nonlocal distortion function. Several recent studies have explored the gravitational consequences of this black-hole solution from different perspectives, including quasinormal behavior and phenomenological constraints \cite{DAgostino:2025sta,Neves:2025uoi,Liu:2025cpp,DAgostino:2026wln,Li:2026zbo,Bravo-Gaete:2026com,DAgostino:2026hht,Sui:2026fjt}.

The first goal of this work is to demonstrate how the single monomial correction can be generalized by superposition. We perturb the Schwarzschild solution by an arbitrary temporal correction \(f(r)\) and solve for the corresponding correction \(h(r)\) to \(\mathcal{B}(r)\), which is the reciprocal of the radial metric component. At first order in the coupling parameter, the reduced field equations are linear in \(f(r)\) and \(h(r)\). As a result, a convergent superposed correction to \(A(r)\) induces a mode-by-mode superposition in \(\mathcal{B}(r)\), while all defining auxiliary fields of the theory can also be obtained by superposing the corresponding single-mode contributions.

The superposition property therefore has a direct physical role. It allows correction terms with different modified-gravity origins to be treated as independent first-order contributions within the same D-W solution-generating structure. In this sense, the known \(r^{-n}\) correction is only the simplest element of a broader class of correction terms that can represent known, scale-dependent, and finite-range effects.

The second goal is to extend the known \(r^{-n}\) correction to other physically motivated asymptotically flat solutions. The logarithmically dressed solution introduces a factor \(\ln(r/\beta)\) in the temporal metric component and decays more slowly than a pure \(r^{-n}\) correction. Such logarithmic terms are especially relevant for nonlocal gravity because they can arise from, quantum effective actions \cite{Barvinsky:1990up}, loop corrections \cite{Donoghue:1994dn}, conformal anomaly effects \cite{Riegert:1984kt}, and nonlocal kernels such as \(R\,\bigl[\ln(-\square/\mu^2)R\bigr]\) \cite{Codello:2015mba,Burzilla:2020bkx}. For the \(n=1\) case, the logarithmic factor can keep the metric noticeably different from the Schwarzschild solution over a wider radial interval. The exponentially suppressed correction introduces the factor \(e^{-r/\beta}\), which can arise from a massive or effective scale \cite{deRham:2014zqa}, a spectral pole \cite{Modesto:2014eta,Giacchini:2016xns}, or screening effects \cite{Khoury:2003aq,Burrage:2017qrf}. Therefore it has the form of a Yukawa-like potential \cite{Capozziello:2009vr}: it describes a finite-range deformation that is suppressed at large radius and mainly affects the near-horizon region. The logarithmic correction behaves in the opposite way. Although it remains asymptotically flat, it slows the return to the Schwarzschild limit.

The structure of this work is as follows. Section~\ref{sec:review} reviews the revised D-W field equations and the black-hole equations written in terms of \(A(r)\), \(\mathcal{B}(r)\), and \(W(r)\). Section~\ref{sec:superposition} establishes the first-order superposition structure and explains the role of \(W'/W\). Section~\ref{sec:other} constructs the logarithmically dressed and exponentially suppressed corrections. Section~\ref{sec:graphical_profiles} compares their qualitative behavior through the temporal metric component and the physical radial metric coefficient. Section~\ref{sec:conclusion} summarizes the main results. Throughout this work we adopt geometric units \(c=\hbar=G=M=1\).

\section{Revised D-W gravity and black hole equations}\label{sec:review}
The action of the revised Deser--Woodard (DW) gravity can be written as \cite{Deser:2019lmm}:
\begin{equation}\label{eq:action}
S=\frac{1}{16\pi}\int \dd^4x\,\sqrt{-g}\,R\bigl(1+F[Y]\bigr),
\end{equation}
where \(F[Y]\) is the nonlocal distortion function. The localized form of the theory can be defined by introducing four auxiliary scalar fields, \(\{X,Y,U,V\}\), satisfying
\begin{align}
\square X &= R, \label{eq:aux_X}\\
\square Y &= g^{\mu\nu}\partial_\mu X \partial_\nu X, \label{eq:aux_Y}\\
\square U &= -2\nabla_\mu\bigl(V\nabla^\mu X\bigr), \label{eq:aux_U}\\
\square V &= R\,F_{,Y}. \label{eq:aux_V}
\end{align}
Here \(\square=g^{\mu\nu}\nabla_\mu\nabla_\nu\). For an arbitrary scalar function \(u\), it is given by
\begin{equation}
\square u=\frac{1}{\sqrt{-g}}\partial_\mu\!\left(\sqrt{-g}\,\partial^\mu u\right).
\end{equation}
Varying the action with respect to the metric gives the vacuum equations of motion
\begin{equation}\label{eq:eom_cov}
\left(G_{\mu\nu}+g_{\mu\nu}\square-\nabla_\mu\nabla_\nu\right)W
+\mathcal{K}_{(\mu\nu)}
-\frac12 g_{\mu\nu}g^{\alpha\beta}\mathcal{K}_{\alpha\beta}=0,
\end{equation}
where $G_{\mu \nu} \equiv R_{\mu \nu}-\frac{1}{2} g_{\mu \nu} R$ is the Einstein tensor, $R_{\mu\nu}$ is the Ricci tensor,
\begin{equation}\label{eq:W_def}
W(r):=1+U(r)+F[Y(r)]
\end{equation}
and
\begin{equation}\label{eq:K_def}
\mathcal{K}_{\mu\nu}:=
\partial_\mu X\partial_\nu U
+\partial_\mu Y\partial_\nu V
+V\partial_\mu X\partial_\nu X .
\end{equation}

The static spherically symmetric line element can be written as:
\begin{equation}\label{eq:metric}
\dd s^2=-A(r)\dd t^2+B(r)\dd r^2
+r^2(\dd\theta^2+\sin^2\theta\,\dd\phi^2),
\end{equation}
where \(B(r)=g_{rr}\). It is convenient to introduce the inverse radial function
\begin{equation}\label{eq:Bcal_def}
\mathcal{B}(r):=\frac{1}{B(r)}.
\end{equation}
The metric is then equivalent to \(\dd s^2=-A(r)\dd t^2+\dd r^2/\mathcal{B}(r)+r^2\dd\Omega^2\). The tetrad frame $\{\mathbf{e}_{\hat t},\mathbf{e}_{\hat r},\mathbf{e}_{\hat \theta},\mathbf{e}_{\hat \phi}\}$ of a
static observer in this spacetime can be related to the tetrad frame $\left\{\mathbf{e}_t, \mathbf{e}_r, \mathbf{e}_\theta, \mathbf{e}_{\phi}\right\}=\left\{\partial_t, \partial_r, \partial_\theta, \partial_{\phi}\right\}$ of a static observer at infinity by:
\begin{equation}
\mathbf{e}_{\hat t}=\frac{\mathbf{e}_t}{\sqrt{A(r)}},
\mathbf{e}_{\hat r}=\sqrt{\mathcal{B}(r)}\,\mathbf{e}_r,
\mathbf{e}_{\hat\theta}=\frac{\mathbf{e}_\theta}{r},
\mathbf{e}_{\hat\phi}=\frac{\mathbf{e}_\phi}{r\sin\theta}.
\end{equation}
In this frame, the two independent components of the field equations eventually take the form~\cite{DAgostino:2025wgl}
\begin{align}
&\left(G_{\hat t\hat t}+G_{\hat\phi\hat\phi}\right)W
=\left(\nabla_{\hat t}\nabla_{\hat t}+\nabla_{\hat\phi}\nabla_{\hat\phi}\right)W,
\label{eq:tetrad_reduced_1}\\
&\left(G_{\hat r\hat r}+G_{\hat\phi\hat\phi}\right)W+2\square W
=\left(\nabla_{\hat r}\nabla_{\hat r}+\nabla_{\hat\phi}\nabla_{\hat\phi}\right)W.
\label{eq:tetrad_reduced_2}
\end{align}
Substituting the metric into these two equations gives~\cite{DAgostino:2025wgl}
    \begin{align}
&\frac{(2A-rA')\left(AW\mathcal{B}'+2A\mathcal{B}W'-\mathcal{B}WA'\right)}{A}
+\frac{4A(\mathcal{B}-1)W}{r}\nonumber\\
&=2r\mathcal{B}WA'', \label{eq:E1}\\
&2\mathcal{B}\left(rWA''+2ArW''+6AW'\right)
+2A\mathcal{B}'(rW'+W)\nonumber\\
&
+A'\left(rW\mathcal{B}'+4\mathcal{B}rW'+6\mathcal{B}W\right)
+\frac{4A(\mathcal{B}-1)W}{r}\nonumber\\
&=\frac{\mathcal{B}rW(A')^2}{A}. \label{eq:E2}
\end{align}
Denote
\begin{equation}\label{eq:F_def}
A_{\rm Sch}(r):=1-\frac{2}{r}.
\end{equation}
Then the Schwarzschild solution corresponds to
\begin{equation}\label{eq:Sch_solution}
A(r)=\mathcal{B}(r)=A_{\rm Sch}(r),\qquad B(r)=\frac{1}{\mathcal{B}(r)}
\end{equation}
The most general form of solution given in Ref.~\cite{DAgostino:2025wgl} is obtained by taking, for \(n>1\),
\begin{equation}\label{eq:singleA}
A(r)=A_{\rm Sch}(r)-\frac{\zeta}{r^n},\qquad
\mathcal{B}(r)=A_{\rm Sch}(r)-\zeta h_n(r).
\end{equation} 

The regular first-order metric correction is
\begin{widetext}
\begin{equation}
h_n(r)
=
\frac{3(r-2)(2r-3)r^n-3^n r\left[n(r-3)(r-2)+4r-9\right]}
{3^n r^{n+1}(r-3)^2}. \label{eq:hn_def}
\end{equation}
\end{widetext}
The corresponding scalar function is
\begin{equation}
W(r)=1+\zeta\,\frac{3^{1-n}-r^{1-n}}{3-r}.
\label{eq:Wn}
\end{equation}
Equivalently, Ref.~\cite{DAgostino:2025wgl} writes \(\mathcal{B}=A_{\rm Sch}+\zeta b_n\) with \(b_n=-h_n\). In the convention of Eq.~\eqref{eq:singleA}, the first cases are
\begin{equation}
\begin{split}
    h_2&=\frac{2r-1}{3r^2},
h_3=\frac{2r^2+5r-9}{9r^3},\\
h_4&=\frac{(2r-3)(r^2+4r+15)}{27r^4}.
\end{split}
\end{equation}
The apparent pole at \(r=3\) is removable, since
\begin{equation}\label{eq:hn3}
\lim_{r\to 3} h_n(r)=\frac{n^2+n+4}{2\,3^{n+1}}.
\end{equation}
In the original derivation, this regularity condition fixes the remaining integration constant.

After solving Eqs.~\eqref{eq:E1}--\eqref{eq:E2}, the auxiliary fields can be reconstructed by the same procedure used in Ref.~\cite{DAgostino:2025wgl}.

\section{Superposition of solutions}\label{sec:superposition}
For a solution with first-order corrections in \(\zeta\), and for an arbitrary single correction function \(f(r)\), we write
\begin{equation}\label{eq:ansatz_general}
A(r)=A_{\rm Sch}(r)-\zeta f(r),\qquad
\mathcal{B}(r)=A_{\rm Sch}(r)-\zeta h(r),
\end{equation}
where \(h(r)\) is determined by the field equations. Substituting Eq.~\eqref{eq:ansatz_general} into Eqs.~\eqref{eq:E1}--\eqref{eq:E2} and retaining only \(O(\zeta)\) terms gives a system that is linear in \(f\), \(h\), and the logarithmic derivative of \(W(r)\). Therefore, once the same boundary and regularity conditions are imposed, the first-order correction to \(\mathcal{B}(r)\) and the corresponding scalar factor \(W(r)\) are fixed linearly by the chosen temporal correction.

An important structural feature of the system is that Eq.~\eqref{eq:E1} determines \(W(r)\) through its logarithmic derivative rather than through \(W\) itself. From Eq.~\eqref{eq:E1} one obtains
\begin{equation}\label{eq:W_log_derivative}
\frac{W'}{W}=\mathcal{I}_W[A,\mathcal{B}],
\end{equation}
with
\begin{equation}\label{eq:IW_general}
\mathcal{I}_W[A,\mathcal{B}]
=
\frac{
2r\mathcal{B}A''
-4A(\mathcal{B}-1)/r
-\frac{2A-rA'}{A}\left(A\mathcal{B}'-\mathcal{B}A'\right)
}{
2\mathcal{B}(2A-rA')
}.
\end{equation}
Expanding around Schwarzschild gives
\begin{equation}
\mathcal{I}_W[A_{\rm Sch}-\zeta f,\,A_{\rm Sch}-\zeta h]
=
\zeta\,\mathcal{I}^{(1)}_W(r;f,h)+\mathcal{O}(\zeta^2),
\end{equation}
where the zeroth-order term vanishes because the Schwarzschild solution has \(W=1\). The coefficient \(\mathcal{I}^{(1)}_W\) is linear in \(f,f',f''\) and in \(h,h'\). Imposing \(W(\infty)=1\), we may therefore write
\begin{equation}\label{eq:W_general_integral}
W(r)
=
\exp\left[
\zeta\int_{\infty}^{r}d\bar r\,
\mathcal{I}^{(1)}_W(\bar r;f,h)
\right].
\end{equation}
Its first-order expansion may
be written as \(W=1+\zeta w+\mathcal{O}(\zeta^2)\), where
\(w'=\mathcal{I}^{(1)}_W\).

Now let the temporal component correction be a convergent sum of sufficiently regular functions,
\begin{equation}
a_\zeta(r)=\sum_n \zeta_n f^{(n)}(r),
\end{equation}
where the superscript \((n)\) labels the correction and is not a derivative. If every single correcting term \(f^{(n)}(r)\) gives a unique first-order metric correction \(h^{(n)}(r)\) under the same boundary and regularity conditions, then linearity gives
\begin{equation}\label{eq:metric_sum_general}
\begin{aligned}
A(r) &= A_{\rm Sch}(r)-\sum_n\zeta_n f^{(n)}(r),\\[2mm]
\mathcal{B}(r) &= A_{\rm Sch}(r)-\sum_n\zeta_n h^{(n)}(r),\\[2mm]
W(r) &= \exp\left[
\int_{\infty}^{r}d\bar r\,
\sum_n\zeta_n\mathcal{I}^{(n)}_W(\bar r)
\right],
\end{aligned}
\end{equation}
where
\begin{equation}
\mathcal{I}^{(n)}_W(r)
\equiv
\mathcal{I}^{(1)}_W(r;f^{(n)},h^{(n)}).
\end{equation}
This solution satisfies the field equations up to products of two or more
amplitudes, denoted collectively by \(O(\zeta_i\zeta_j)\). Expanding the
exponential gives the equivalent first-order expression
\begin{equation}
W(r)=1+\sum_n\zeta_n w^{(n)}(r)+O(\zeta_i\zeta_j).
\end{equation}

The result of Ref.~\cite{DAgostino:2025wgl} is recovered by choosing
\begin{equation}
f^{(n)}(r)=r^{-n},\qquad n>1,
\end{equation}
for which \(h^{(n)}(r)=h_n(r)\) is given in Eq.~\eqref{eq:hn_def}, while the corresponding scalar factor is Eq.~\eqref{eq:Wn}. The same first-order structure also applies to other asymptotically flat correction functions, including the logarithmically dressed and exponentially suppressed correction terms discussed below, for example
\begin{equation}
f^{(n)}_{\log}(r)=\frac{\ln(r/\beta)}{r^n},
\qquad
f^{(n)}_{\exp}(r)=-\frac{e^{-r/\beta}}{r^n},
\end{equation}
where the sign in the exponential case follows the convention
\(A=A_{\rm Sch}-\zeta f\) and may equivalently be absorbed into \(\zeta\).

This observation also gives a physical interpretation of the superposition
structure. The functions \(f^{(n)}(r)\) can be regarded as elementary correction
terms associated with different contributions to a modified-gravity solution.
The \(r^{-n}\) corrections give the known corrections
of the revised D-W solution, logarithmically dressed corrections represent
scale-dependent effects, and exponentially suppressed corrections represent
finite-range or screened effects. The linearity of the first-order D-W
equations then promotes these terms into a consistent set of superposable
correction terms for \(A(r)\), \(\mathcal{B}(r)\), and \(W(r)\).

Interestingly, this linearity is a property about the inverse radial metric function
\(\mathcal{B}(r)\), not about the physical radial component
\(B(r)=g_{rr}=1/\mathcal{B}(r)\). Defining
\begin{equation}
H_\zeta(r):=\sum_n\zeta_n h^{(n)}(r),
\end{equation}
we have
\begin{equation}\label{eq:grr_resummed}
B(r)
=
\frac{1}{A_{\rm Sch}(r)-H_\zeta(r)}
=
\frac{1}{A_{\rm Sch}(r)}
+\frac{H_\zeta(r)}{A_{\rm Sch}(r)^2}
+\frac{H_\zeta(r)^2}{A_{\rm Sch}(r)^3}
+\cdots .
\end{equation}
Thus \(\mathcal{B}(r)\) is linear at first order, while \(B(r)\) can contain correction contributions that are higher-order terms in \(\zeta\).

\section{Extended spherically symmetric solutions}\label{sec:other}
In this section, we discuss two cases that can extend the static spherically symmetric solutions of D-W nonlocal gravity: a logarithmically dressed correction and an exponentially suppressed correction.

The two correction terms are chosen because the same mathematical structures
are allowed by well-known physical mechanisms in modified gravity. The
logarithmically dressed correction contains an explicit scale \(\beta\), which
is characteristic of scale-dependent effective gravitational dynamics. The
exponentially suppressed correction contains a finite range controlled by
\(\beta\), which is characteristic of massive scales, spectral poles, or
screening lengths.

\subsection{Logarithmically dressed correction}
\label{subsec:log_correction}

Let us now search for a new class of asymptotically flat black-hole solutions for D-W nonlocal gravity.
Instead of a pure \(r^{-n}\) correction, we consider a logarithmically
dressed correction to the temporal metric component as the initial ansatz. In this subsection, the
logarithm is imposed with an arbitrary reference scale \(\beta\), and we write
\begin{equation}
    \ell_r\equiv \ln\frac{r}{\beta},
    \qquad
    L_\beta\equiv \ln\frac{3}{\beta}.
    \label{log_defs}
\end{equation}
The general-\(n\) ansatz is
\begin{equation}
    A(r)=1-\frac{2}{r}
    -\zeta\,\frac{\ell_r}{r^n},
    \label{A_log}
\end{equation}
where \(n\in\mathbb{N}\), \(|\zeta|\ll1\), and \(\beta>0\). The table below then
specializes the resulting auxiliary-field solution to \(n=1,2,3\).

The logarithmic factor has a direct physical meaning. In effective descriptions
of modified gravity, logarithmic terms can appear from quantum effective
actions, loop corrections, renormalization-group running, conformal anomaly
effects, and massless nonlocal kernels such as \(R\,\bigl[\ln(-\square/\mu^2)R\bigr]\). In
the present solution, \(\beta\) plays the role of the reference scale, and the
factor \(\ln(r/\beta)\) can be interpreted as an infrared or scale-running
correction to the Schwarzschild geometry.

As in the previous case, we perturb the inverse radial metric function as
\begin{equation}
    \mathcal{B}(r)=1-\frac{2}{r}-\zeta h(r).
    \label{B_log_ansatz}
\end{equation}
Substituting Eqs.~\eqref{A_log} and \eqref{B_log_ansatz} into Eq.~\eqref{eq:E1} gives the formal integral solution
\begin{equation}
    W(r)
    =
    w_1
    \exp\left[
    \int dr\,\mathcal{I}_{W,\log}(r)
    \right],
    \label{W_log_integral}
\end{equation}
where \(w_1\) is an integration constant. The integrand can be written as
\begin{equation}
    \mathcal{I}_{W,\log}(r)
    =
    \frac{
    \zeta\,\mathcal{N}_{W,\log}(r)
    }{
    (r-2)\mathcal{D}_{W,\log}(r)
    },
    \label{IW_log}
\end{equation}
with
\begin{equation}
\begin{split}
    \mathcal{N}_{W,\log}(r)
    &=
    2-2n(r-2)^2-r
    -2r^n\left[3+r(r-3)\right]h(r)
    \\
    &\quad
    +\left[
    n\left(1+n(r-2)\right)(r-2)+2r-2
    \right]\ell_r
    \\
    &\quad
    -(r-3)(r-2)r^{n+1}h'(r),
\end{split}
\label{NW_log}
\end{equation}
and
\begin{equation}
\begin{split}
    \mathcal{D}_{W,\log}(r)
    &=
    2(r-3)r^n\left[2-r+r\zeta h(r)\right]
    \\
    &\quad
    +(r-2)r\zeta
    \left[-1+(n+2)\ell_r\right].
\end{split}
\label{DW_log}
\end{equation}

Substituting Eq.~\eqref{W_log_integral} into the second independent field
equation and solving at first order in \(\zeta\), we find
\begin{equation}
\begin{split}
    h(r)
    &=
    \frac{(r-2)(2r-3)}{r(r-3)^2}
    \bigg[
    \frac{(r-3)r^{1-n}}{2r-3}
    +3^{1-n}L_\beta
    \\
    &\hspace{0.9cm}
    -
    \frac{
    r^{1-n}
    \left[
    -9+6n+(4-5n)r+nr^2
    \right]\ell_r
    }
    {(r-2)(2r-3)}
    \bigg].
\end{split}
\label{h_log}
\end{equation}
Equivalently, the inverse radial metric function can be written as
\begin{equation}
\begin{split}
    \mathcal{B}(r)
    &=
    1-\frac{2}{r}
    +\frac{\zeta}{(r-3)^2}
    \bigg[
    \frac{3^{1-n}(r-2)(3-2r)}{r}L_\beta
    \\
    &\quad
    +r^{-n}
    \bigg\{
    -(r-3)(r-2)
    +\left[n(r-3)(r-2)+4r-9\right]\ell_r
    \bigg\}
    \bigg].
\end{split}
\label{B_log}
\end{equation}
Although Eqs.~\eqref{h_log} and \eqref{B_log} seem to contain a pole at
\(r=3\), this singularity is removable. Indeed,
\begin{equation}
    \lim_{r\rightarrow 3}h(r)
    =
    -1-2n+(4+n+n^2)L_\beta .
    \label{h_log_limit}
\end{equation}

After the regular solution \eqref{h_log} is inserted and the result is expanded
consistently to first order in \(\zeta\), the integrand becomes
\begin{equation}
    \mathcal{I}_{W,\log}(r)
    =
    \zeta\frac{d}{dr}
    \left[
    \frac{r^{1-n}\ell_r-3^{1-n}L_\beta}{r-3}
    \right]
    +\mathcal{O}(\zeta^2).
    \label{IW_log_integrated}
\end{equation}
With the integration constant chosen so as to recover the Schwarzschild limit at
spatial infinity, the scalar function \(W(r)\) takes the simple first-order form
\begin{equation}
    W(r)=1+\zeta\,\omega_n(r),
    \label{W_log_final}
\end{equation}
where
\begin{equation}
    \omega_n(r)
    =
    \frac{r^{1-n}\ell_r-3^{1-n}L_\beta}{r-3}.
    \label{omega_log}
\end{equation}
The apparent pole at \(r=3\) is again removable.

The event horizon is determined by \(A(r_H)=0\). For the general logarithmic
scale one finds
\begin{equation}
    r_H=2+\zeta\,2^{1-n}\ln\frac{2}{\beta}+\mathcal{O}(\zeta^2).
    \label{horizon_log}
\end{equation}
The special normalization \(\beta=2\) therefore removes the first-order horizon
shift, but this cancellation is not present for a generic reference scale.

The Ricci scalar associated with Eqs.~\eqref{A_log} and \eqref{B_log} is
\begin{equation}
\begin{split}
    &R(r)\\
    &=
    \frac{3\zeta r^{-n-2}}{(r-3)^3}
    \bigg[
    -(r-3)
    \left[-12+2n(r-3)(r-2)-(r-9)r\right]
    \\
    &\quad
    -2\,3^{1-n}r^n(2r-3)L_\beta+6(2r-3)\ell_r
    \\
    &\quad
    +\left\{n(r-3)
    \left[-12+n(r-3)(r-2)-(r-9)r\right]
    \right\}\ell_r
    \bigg].
\end{split}
\label{Ricci_log}
\end{equation}
At \(r=3\), this gives the finite limit
\begin{equation}
\begin{split}
    \lim_{r\rightarrow 3}R(r)
    & =
    3^{-n-3}\zeta
    \bigg[
    5-12n+3n^2
    \\
    &\quad
    +\left(-n^3+6n^2-5n\right)L_\beta
    \bigg].
\end{split}
\label{Ricci_log_limit}
\end{equation}

We now reconstruct the auxiliary fields. Solving \(\square X=R\), one obtains
\begin{equation}
    X(r)
    =
    \zeta
    \left[
    \frac{3r^{1-n}\ell_r-3^{2-n}L_\beta}{r-3}
    +x_1\ln\left(\frac{r}{r-2}\right)
    \right],
    \label{X_log}
\end{equation}
where \(x_1\) is an integration constant. The second auxiliary field is
\begin{equation}
    Y(r)=\zeta\ln\left(\frac{r}{r-2}\right).
    \label{Y_log}
\end{equation}
We can also choose
\begin{equation}
    U(r)=0.
    \label{U_log_sector}
\end{equation}
The equation for \(V(r)\) is linear but its general-\(n\) primitive is lengthy;
therefore the first three explicit cases are listed in
Table~\ref{tab:log_solutions_compact}.

Since \(W=1+U+F[Y]\), the radial representation of the distortion function is
\begin{equation}
    F[Y(r)]
    =
    \zeta
    \frac{r^{1-n}\ell_r-3^{1-n}L_\beta}{r-3}.
    \label{F_log_r}
\end{equation}
Inverting Eq.~\eqref{Y_log}, we find
\begin{equation}
    r(Y)=\frac{2e^{Y/\zeta}}{e^{Y/\zeta}-1}.
    \label{r_of_Y_log}
\end{equation}
Thus the distortion function can be written as
\begin{equation}
\begin{split}
    F[Y]
    &=
    \zeta
    \frac{e^{Y/\zeta}-1}{3-e^{Y/\zeta}}
    \bigg[
    \left(
    \frac{2e^{Y/\zeta}}{e^{Y/\zeta}-1}
    \right)^{1-n}
    \\
    &\hspace{1.2cm}
    \times
    \ln\left(
    \frac{2e^{Y/\zeta}}{\beta\left(e^{Y/\zeta}-1\right)}
    \right)
    -3^{1-n}L_\beta
    \bigg].
\end{split}
\label{F_log_Y}
\end{equation}
The Schwarzschild limit is recovered asymptotically, since
\(F[Y]\rightarrow0\) as \(r\rightarrow\infty\), or equivalently \(Y\rightarrow0\).

For compactness in the table, we also use
\begin{equation}
    q\equiv \ell_r-L_\beta=\ln\frac{r}{3},
    \qquad
    G_3\equiv \ln 3,
    \qquad
    E\equiv e^{Y/\zeta}.
\end{equation}
For the reconstructed \(V(r)\) entries, we use the compact primitive basis
\(\mathcal{J}_{k,a,s}\). To keep the exterior-region expression explicitly real,
we use the real-valued convention
\begin{equation}
    \mathcal{L}_a(r)\equiv\ln\left|1-\frac{r}{a}\right|,
    \qquad
    \operatorname{Li}^{\rm R}_m(x)\equiv
    \operatorname{Re}\!\left[\operatorname{Li}_m(x)\right],
\end{equation}
where $\operatorname{Li}_m(x)$ are the Polylogarithms. The basis is then defined for \(k=0,1,2\) by
\begin{widetext}
\begin{equation}
\begin{alignedat}{2}
\mathcal{J}_{k,0,1}
&=\frac{q^{k+1}}{k+1},
&\qquad \mathcal{J}_{k,0,s}
&=-r^{1-s}\sum_{j=0}^{k}
\frac{k!}{(k-j)!}\frac{q^{k-j}}{(s-1)^{j+1}},
\quad s>1,
\\[1mm]
\mathcal{J}_{0,a,1}
&=\mathcal{L}_a(r),
&\qquad \mathcal{J}_{1,a,1}
&=q\mathcal{L}_a(r)
+\operatorname{Li}^{\rm R}_2\left(\frac{r}{a}\right),
\quad a\ne0,
\\[1mm]
\mathcal{J}_{2,a,1}
&=q^2\mathcal{L}_a(r)
+2q\operatorname{Li}^{\rm R}_2\left(\frac{r}{a}\right)
-2\operatorname{Li}^{\rm R}_3\left(\frac{r}{a}\right),
&\qquad \mathcal{J}_{0,a,s}
&=-\frac{1}{(s-1)(r-a)^{s-1}},
\quad a\ne0,\ s>1,
\\[1mm]
\mathcal{J}_{k,a,s}
&\mathrlap{=
-\frac{q^k}{(s-1)(r-a)^{s-1}}
+\frac{k}{s-1}
\left[
\frac{(-1)^{s-1}}{a^{s-1}}\mathcal{J}_{k-1,0,1}
+\sum_{j=1}^{s-1}
\frac{(-1)^{s-1-j}}{a^{s-j}}\mathcal{J}_{k-1,a,j}
\right],
\quad a\ne0,\ s>1,\ k=1,2,}
& &
\end{alignedat}
\label{J_log_basis}
\end{equation}
\end{widetext}
The real-valued form differs from the corresponding principal-valued primitive
only by constant terms, which can be absorbed into the homogeneous constants
of \(V(r)\). All of the corresponding quantities for the \(n=1,2,3\) cases are summarized in
Table~\ref{tab:log_solutions_compact}.

\begin{table*}[t]
\centering
\fontsize{4.6}{5.5}\selectfont
\setlength{\tabcolsep}{1.5pt}
\renewcommand{\arraystretch}{2.0}
\newcommand{\mcell}[1]{\(\begin{aligned}[t]#1\end{aligned}\)}
\newcommand{\J}[3]{\mathcal{J}_{#1,#2,#3}}
\newcommand{\Lb}{L_\beta}
\newcommand{\Gthree}{G_3}
\begin{tabular}{@{}p{0.10\textwidth}p{0.29\textwidth}p{0.29\textwidth}p{0.29\textwidth}@{}}
\hline
&
\(n=1\)
&
\(n=2\)
&
\(n=3\)
\\
\hline

\(\begin{array}{c}A(r)\\ \mathcal{B}(r)\end{array}\)
&
\mcell{
A(r)&=1-\frac{2}{r}-\zeta\frac{\ell_r}{r},
\\[1mm]
\mathcal{B}(r)&=1-\frac{2}{r}
+\frac{\zeta}{r(r-3)^2}
\Big[
(r-2)(3-2r)L_\beta
\\
&\quad
-(r-3)(r-2)
+\left(r^2-r-3\right)\ell_r
\Big].
}
&
\mcell{
A(r)&=1-\frac{2}{r}-\zeta\frac{\ell_r}{r^2},
\\[1mm]
\mathcal{B}(r)&=1-\frac{2}{r}
+\frac{\zeta}{3r^2(r-3)^2}
\Big[
r(r-2)(3-2r)L_\beta
\\
&\quad
-3(r-3)(r-2)
+3\left(2r^2-6r+3\right)\ell_r
\Big].
}
&
\mcell{
A(r)&=1-\frac{2}{r}-\zeta\frac{\ell_r}{r^3},
\\[1mm]
\mathcal{B}(r)&=1-\frac{2}{r}
+\frac{\zeta}{9r^3(r-3)^2}
\Big[
r^2(r-2)(3-2r)L_\beta
\\
&\quad
-9(r-3)(r-2)
+9\left(3r^2-11r+9\right)\ell_r
\Big].
}
\\
\hline

\(\begin{array}{c}X(r)\\ Y(r)\end{array}\)
&
\mcell{
X(r)&=\zeta\left[
\frac{3(\ell_r-L_\beta)}{r-3}
+x_1\ln\left(\frac{r}{r-2}\right)
\right],
\\[1mm]
Y(r)&=\zeta\ln\left(\frac{r}{r-2}\right).
}
&
\mcell{
X(r)&=\zeta\left[
\frac{3\ell_r-rL_\beta}{r(r-3)}
+x_1\ln\left(\frac{r}{r-2}\right)
\right],
\\[1mm]
Y(r)&=\zeta\ln\left(\frac{r}{r-2}\right).
}
&
\mcell{
X(r)&=\zeta\left[
\frac{9\ell_r-r^2L_\beta}{3r^2(r-3)}
+x_1\ln\left(\frac{r}{r-2}\right)
\right],
\\[1mm]
Y(r)&=\zeta\ln\left(\frac{r}{r-2}\right).
}\\
\ 
\\
\hline

\(\begin{array}{c}U(r)\\ V(r)\end{array}\)
&
\mcell{
U(r)&=0,
\\
V(r)&=v_0+v_1\ln\frac{r}{r-2}
-\frac{3}{4}\J{2}{3}{2}
-3\J{2}{3}{3}
\\
&\quad
-\frac{9}{4}\J{2}{3}{4}
+\frac{3}{2}\J{1}{3}{2}
+\frac{3}{2}\J{1}{3}{3}
\\
&\quad
+\frac{3}{8}(\Gthree+1)^2\J{0}{2}{1}
\\
&\quad
-\frac{1}{6}\J{0}{3}{1}
\\
&\quad
-\frac{1}{4}\J{0}{3}{2}
-\frac{(3\Gthree+1)(3\Gthree+5)}{24}
\J{0}{0}{1}.
}
&
\mcell{
U(r)&=0,
\\
V(r)&=v_0+v_1\ln\frac{r}{r-2}
+\frac{1}{27}\J{2}{3}{1}
-\frac{1}{36}\J{2}{3}{2}
\\
&\quad
-\frac{1}{3}\J{2}{3}{3}
-\frac{1}{4}\J{2}{3}{4}
-\frac{1}{27}\J{2}{0}{1}
\\
&\quad
-\frac{1}{12}\J{2}{0}{2}
+\frac{1}{6}\J{2}{0}{3}
+\frac{2\Lb-3}{54}\J{1}{3}{1}
\\
&\quad
+\frac{\Lb+2}{18}\J{1}{3}{2}
+\frac{1}{6}\J{1}{3}{3}
-\frac{2\Lb-3}{54}\J{1}{0}{1}
\\
&\quad
-\frac{3\Lb-1}{18}\J{1}{0}{2}
+\frac{\Lb-1}{3}\J{1}{0}{3}
+\frac{\Lb^2}{24}\J{0}{2}{1}
\\
&\quad
-\frac{\Lb}{54}\J{0}{3}{1}
-\frac{1}{36}\J{0}{3}{2}
-\frac{\Lb(9\Lb-4)}{216}\J{0}{0}{1}
\\
&\quad
-\frac{(\Lb-1)(3\Lb+1)}{36}
\J{0}{0}{2}
+\frac{(\Lb-1)^2}{6}\J{0}{0}{3}.
}
&
\mcell{
U(r)&=0,
\\
V(r)&=v_0+v_1\ln\frac{r}{r-2}
+\frac{2}{243}\J{2}{3}{1}
+\frac{1}{108}\J{2}{3}{2}
\\
&\quad
-\frac{1}{27}\J{2}{3}{3}
-\frac{1}{36}\J{2}{3}{4}
-\frac{2}{243}\J{2}{0}{1}
\\
&\quad
-\frac{11}{324}\J{2}{0}{2}
-\frac{5}{54}\J{2}{0}{3}
-\frac{1}{9}\J{2}{0}{4}
+\frac{2}{3}\J{2}{0}{5}
\\
&\quad
+\frac{2\Lb-3}{243}\J{1}{3}{1}
+\frac{3\Lb+1}{162}\J{1}{3}{2}
+\frac{1}{54}\J{1}{3}{3}
\\
&\quad
-\frac{2\Lb-3}{243}\J{1}{0}{1}
-\frac{7\Lb-5}{162}\J{1}{0}{2}
-\frac{10\Lb-3}{54}\J{1}{0}{3}
\\
&\quad
-\frac{2\Lb}{9}\J{1}{0}{4}
+\frac{2(2\Lb-1)}{3}\J{1}{0}{5}
+\frac{\Lb^2}{216}\J{0}{2}{1}
\\
&\quad
-\frac{3\Lb-1}{486}\J{0}{3}{1}
-\frac{1}{324}\J{0}{3}{2}
-\frac{(3\Lb-2)^2}{1944}\J{0}{0}{1}
\\
&\quad
-\frac{3\Lb^2-6\Lb+1}{324}
\J{0}{0}{2}
-\frac{\Lb(5\Lb-3)}{54}\J{0}{0}{3}
\\
&\quad
-\frac{(2\Lb-1)(2\Lb+1)}{36}
\J{0}{0}{4}
+\frac{(2\Lb-1)^2}{6}\J{0}{0}{5}.
}
\\
\ 
\\
\hline

\(F[Y]\)
&
\mcell{
F[Y]&=
\zeta\frac{E-1}{3-E}
\left[
\ln\left(
\frac{2E}{\beta(E-1)}
\right)
-L_\beta
\right].
}
&
\mcell{
F[Y]&=
\zeta\frac{E-1}{3-E}
\left[
\frac{E-1}{2E}
\ln\left(
\frac{2E}{\beta(E-1)}
\right)
-\frac{L_\beta}{3}
\right].
}
&
\mcell{
F[Y]&=
\zeta\frac{E-1}{3-E}
\left[
\frac{(E-1)^2}{4E^2}
\ln\left(
\frac{2E}{\beta(E-1)}
\right)
-\frac{L_\beta}{9}
\right].
}
\\
\hline
\end{tabular}
\caption{
Summary of the logarithmically corrected black hole solutions for
\(n=1,2,3\), obtained by specializing the general-\(n\) logarithmic correction. We use \(\ell_r\equiv\ln(r/\beta)\), \(L_\beta\equiv\ln(3/\beta)\),
\(q\equiv\ell_r-L_\beta\), \(G_3\equiv\ln3\), and \(E\equiv e^{Y/\zeta}\).
The basis \(\mathcal{J}_{k,a,s}\) is defined in Eq.~\eqref{J_log_basis} using
the real-valued notation \(\mathcal{L}_a(r)=\ln|1-r/a|\) and
\(\operatorname{Li}^{\rm R}_m(x)=\operatorname{Re}[\operatorname{Li}_m(x)]\).
The constants
\(v_0\) and \(v_1\) are the homogeneous integration constants in the
reconstructed \(V(r)\).
}
\label{tab:log_solutions_compact}
\end{table*}

\subsection{Exponentially suppressed correction}
\label{subsec:exp_correction}

Let us further consider an exponentially suppressed correction to the
Schwarzschild temporal metric component. We take
\begin{equation}
    A(r)=1-\frac{2}{r}
    +\zeta\,\frac{e^{-r/\beta}}{r^n},
    \label{A_exp}
\end{equation}
where \(n\in\mathbb{N}\), \(|\zeta|\ll1\), and \(\beta>0\). The parameter
\(\beta\) controls the range of the correction. Therefore, unlike a pure \(r^{-n}\) correction, the correction terms in this case are exponentially suppressed at large radius and are
automatically asymptotically flat.

The exponential factor also has a direct physical meaning. In many modified
gravity scenarios, a finite correlation length, a spectral pole, a screening
length, or an effective massive scale produces Yukawa-like correction terms. The
parameter \(\beta\) then controls the distance over which the correction remains
active. In the present D-W solution, this makes the correction naturally
concentrated near the black-hole region while keeping the large-radius geometry
close to the Schwarzschild limit.

As in the previous cases, we perturb the inverse radial metric function,
\begin{equation}
    \mathcal{B}(r)=1-\frac{2}{r}-\zeta h(r).
    \label{B_exp_ansatz}
\end{equation}
Substituting Eqs.~\eqref{A_exp} and \eqref{B_exp_ansatz} into Eq.~\eqref{eq:E1} gives the formal integral solution
\begin{equation}
    W(r)
    =
    w_1
    \exp\left[
    \int dr\,\mathcal{I}_W(r)
    \right],
    \label{W_exp_integral}
\end{equation}
where the integral is indefinite and \(w_1\) is an integration constant. The integrand can be written as
\begin{equation}
    \mathcal{I}_W(r)
    =
    \frac{
    \zeta\,\mathcal{N}_W(r)
    }
    {
    (r-2)\beta\,\mathcal{D}_W(r)
    },
    \label{IW_exp}
\end{equation}
with
\begin{equation}
\begin{split}
    \mathcal{N}_W(r)
    &=
    r^2(r-2)^2
    \\
    &\quad
    +r\left[
    -6+8n+(5-8n)r+(2n-1)r^2
    \right]\beta
    \\
    &\quad
    +\left[
    4n^2-2n-2+(n+2-4n^2)r+n^2r^2
    \right]\beta^2
    \\
    &\quad
    +2e^{r/\beta}r^n
    \left(r^2-3r+3\right)\beta^2 h(r)
    \\
    &\quad
    +e^{r/\beta}r^{n+1}
    (r-2)(r-3)\beta^2 h'(r),
\end{split}
\label{NW_exp}
\end{equation}
and
\begin{equation}
\begin{split}
    \mathcal{D}_W(r)
    &=
    2e^{r/\beta}r^n\beta(r-2)(r-3)
    \\
    &\quad
    +\zeta r(r-2)\left[r+(n+2)\beta\right]
    \\
    &\quad
    -2\zeta\beta e^{r/\beta}r^{n+1}(r-3)h(r).
\end{split}
\label{DW_exp}
\end{equation}

Substituting Eq.~\eqref{W_exp_integral} into the second independent field
equation~\eqref{eq:E2} and solving at first order in \(\zeta\), we find
\begin{equation}
\begin{split}
    h(r)
    &=
    \frac{1}{(r-3)^2}
    \bigg[
    -\frac{
    3^{1-n}e^{-3/\beta}(r-2)(2r-3)
    }{r}
    \\
    &\quad
    +\frac{
    e^{-r/\beta}r^{-n}
    }{\beta}
    \bigg\{
    r(r-3)(r-2)
    \\
    &\qquad
    +\beta\left[
    n(r-3)(r-2)+4r-9
    \right]
    \bigg\}
    \bigg].
\end{split}
\label{h_exp}
\end{equation}
Equivalently, the inverse radial metric function is
\begin{equation}
\begin{split}
    \mathcal{B}(r)
    &=
    1-\frac{2}{r}
    +\frac{
    \zeta\,3^{1-n}e^{-3/\beta}(r-2)(2r-3)
    }{
    r(r-3)^2
    }
    \\
    &\quad
    -\frac{
    \zeta e^{-r/\beta}r^{-n}
    }{
    \beta(r-3)^2
    }
    \bigg\{
    r(r-3)(r-2)
    \\
    &\qquad
    +\beta\left[
    n(r-3)(r-2)+4r-9
    \right]
    \bigg\}.
\end{split}
\label{B_exp}
\end{equation}

Although Eqs.~\eqref{h_exp} and \eqref{B_exp} contain powers of
\((r-3)^{-1}\), the singularity at \(r=3\) is removable. Indeed,
\begin{equation}
    \lim_{r\rightarrow3}h(r)
    =
    \frac{
    3^{-1-n}e^{-3/\beta}
    \left[
    -9-6n\beta-(4+n+n^2)\beta^2
    \right]
    }{
    \beta^2
    }.
    \label{h_exp_limit}
\end{equation}
Therefore, as in the previous cases, the point \(r=3\) does not represent an
essential singularity of the geometry.

As in the previous solutions, we fix the remaining homogeneous
constant by requiring \(W(r)\rightarrow1\) as \(r\rightarrow\infty\). The
asymptotically normalized expression is
\begin{equation}
    W(r)=1+\zeta\,\omega_n(r),
    \label{W_exp_final}
\end{equation}
where
\begin{equation}
    \omega_n(r)
    =
    \frac{
    3^{1-n}e^{-3/\beta}
    -e^{-r/\beta}r^{1-n}
    }{
    r-3
    }.
    \label{omega_exp}
\end{equation}
The pole at \(r=3\) is again removable.

The event horizon is determined by \(A(r_H)=0\). Writing
\(r_H=2+\zeta\,\delta r_H+\mathcal{O}(\zeta^2)\), one obtains
\begin{equation}
    r_H
    =
    2-\frac{\zeta e^{-2/\beta}}{2^{n-1}}
    +\mathcal{O}(\zeta^2).
    \label{horizon_exp}
\end{equation}
Hence, for positive \(\zeta\), the exponentially suppressed correction shifts the
horizon inward at first order.

The Ricci scalar corresponding to Eqs.~\eqref{A_exp} and \eqref{B_exp} is
\begin{equation}
\begin{split}
    &R(r)\\
    &=
    \frac{
    \zeta e^{-(3+r)/\beta}r^{-n-2}
    }{
    (r-3)^3\beta^2
    }
    \bigg[
    2\,3^{2-n}e^{r/\beta}r^n(2r-3)\beta^2
    \\
    &
    -3e^{3/\beta}
    \bigg\{
    (r-3)^2(r-2)r^2
    \\
    &
    +2(r-3)r
    \left[
    -9+n(r-3)(r-2)-r(r-7)
    \right]\beta
    \\
    &
    +
    \bigg[
    6(2r-3)
    \\
    &
    +n(r-3)
    \left[
    -12+n(r-3)(r-2)-r(r-9)
    \right]
    \bigg]\beta^2
    \bigg\}
    \bigg].
\end{split}
\label{Ricci_exp}
\end{equation}
At \(r=3\), this admits the finite limit
\begin{equation}
\begin{split}
    \lim_{r\rightarrow3}R(r)
    &=
    \frac{
    3^{-3-n}e^{-3/\beta}\zeta
    }{
    \beta^3
    }
    \bigg[
    27+27(n-3)\beta
    \\
    &\quad
    +9(n^2-5n+4)\beta^2
    +n(n^2-6n+5)\beta^3
    \bigg].
\end{split}
\label{Ricci_exp_limit}
\end{equation}

We now reconstruct the auxiliary fields. Solving \(\square X=R\), and choosing
the integration constant such that \(X(r)\rightarrow0\) at infinity, one obtains
\begin{equation}
    X(r)
    =
    \zeta
    \left[
    \frac{
    3\left(
    3^{1-n}e^{-3/\beta}
    -e^{-r/\beta}r^{1-n}
    \right)
    }{
    r-3
    }
    +x_1\ln\left(\frac{r}{r-2}\right)
    \right],
    \label{X_exp}
\end{equation}
where \(x_1\) is an integration constant associated with the homogeneous mode.
The second auxiliary field is again
\begin{equation}
    Y(r)=\zeta\ln\left(\frac{r}{r-2}\right).
    \label{Y_exp}
\end{equation}
We also take
\begin{equation}
    U(r)=0.
    \label{U_exp_sector}
\end{equation}
The corresponding reconstructed \(V(r)\) is again very lengthy, so the first three explicit
cases are displayed in Table~\ref{tab:exp_solutions_compact}.

Since \(W=1+U+F[Y]\) and \(U=0\), the radial representation of the distortion function is
\begin{equation}
    F[Y(r)]
    =
    \zeta
    \frac{
    3^{1-n}e^{-3/\beta}
    -e^{-r/\beta}r^{1-n}
    }{
    r-3
    }.
    \label{F_exp_r}
\end{equation}
Inverting Eq.~\eqref{Y_exp}, we find
\begin{equation}
    r(Y)=\frac{2e^{Y/\zeta}}{e^{Y/\zeta}-1}.
    \label{r_of_Y_exp}
\end{equation}
Therefore, defining
\begin{equation}
    E\equiv e^{Y/\zeta},
    \qquad
    \mathcal{R}(Y)\equiv \frac{2E}{E-1},
    \label{ER_def_exp}
\end{equation}
the distortion function may be written compactly as
\begin{equation}
    F[Y]
    =
    \frac{\zeta}{\mathcal{R}(Y)-3}
    \left[
    3^{1-n}e^{-3/\beta}
    -e^{-\mathcal{R}(Y)/\beta}\mathcal{R}(Y)^{1-n}
    \right].
    \label{F_exp_Y}
\end{equation}
By construction, \(F[Y]\rightarrow0\) as \(r\rightarrow\infty\), or equivalently
as \(Y\rightarrow0\).

For compactness, in Table~\ref{tab:exp_solutions_compact} we use
\begin{equation}
    \mathcal{E}_r\equiv e^{-r/\beta},
    \qquad
    \mathcal{E}_3\equiv e^{-3/\beta},
    \qquad
    \mathcal{R}\equiv \frac{2E}{E-1},
    \qquad
    E\equiv e^{Y/\zeta}.
\end{equation}
For the \(V(r)\) entries we additionally define
\begin{widetext}
\begin{equation}
\begin{alignedat}{3}
\Delta&\equiv r-3,
&\hspace{1em}\Ei_1&\equiv \operatorname{Ei}\!\left(\frac{3-r}{\beta}\right),
&\hspace{1em}\Ei_2&\equiv \operatorname{Ei}\!\left(\frac{6-2r}{\beta}\right),
\\[1mm]
\Ei_r&\equiv \operatorname{Ei}\!\left(-\frac{r}{\beta}\right),
&\hspace{1em}\Ei_{2r}&\equiv \operatorname{Ei}\!\left(-\frac{2r}{\beta}\right),
&\hspace{1em}A_\beta&\equiv -9+9\beta+6\beta^2+2\beta^3,
\\[1mm]
B_\beta&\equiv 9-6\beta+2\beta^2,
&\hspace{1em}C_\beta&\equiv -9+6\beta^2+4\beta^3,
&\hspace{1em}D_\beta&\equiv 54-45\beta+36\beta^2-18\beta^3+4\beta^4,
\\[1mm]
P_2&\equiv -3\Delta^2+3\Delta(r-2)\beta+(3+2\Delta r)\beta^2,
&\hspace{1em}Q_2&\mathrlap{\equiv -3\Delta^2r^2+6\Delta r(4r-9)\beta
+\big(81-162r+102r^2-30r^3+4r^4\big)\beta^2,}
&&
\\[1mm]
P_3&\mathrlap{\equiv -3\Delta^2r+3\Delta r\beta
+\big(-108+93r-30r^2+4r^3\big)\beta^2,}
&&&&
\\[1mm]
Q_3&\mathrlap{\equiv 18\Delta^3r^3-3\Delta^2r^2(-9+4\Delta r)\beta
+6\Delta r\big[54-18r+r^2(r-4)(2r-3)\big]\beta^2}
&&&&
\\
&\mathrlap{\quad
+\big[-1458+1782r-405r^2+54r^3-132r^4+60r^5-8r^6\big]\beta^3,}
&&&&
\end{alignedat}
\label{V_exp_notation}
\end{equation}
\end{widetext}
where \(\operatorname{Ei}\) is the exponential integral function. Although the
compact expressions for \(V(r)\) contain explicit powers of \(\Delta^{-1}\) and
exponential-integral terms, their apparent singular parts cancel at \(r=3\) for
the listed \(n=1,2,3\) cases. This cancellation has been verified by expanding
the full expressions symbolically about \(r=3\). The first three cases are
listed in Table~\ref{tab:exp_solutions_compact}.

\begin{table*}[t]
\centering
\fontsize{4.6}{5.5}\selectfont
\setlength{\tabcolsep}{1.5pt}
\renewcommand{\arraystretch}{2.15}
\newcommand{\mcell}[1]{\(\begin{aligned}[t]#1\end{aligned}\)}
\begin{tabular}{@{}p{0.10\textwidth}p{0.29\textwidth}p{0.29\textwidth}p{0.29\textwidth}@{}}
\hline
&
\(n=1\)
&
\(n=2\)
&
\(n=3\)
\\
\hline

\(\begin{array}{c}A(r)\\ \mathcal{B}(r)\end{array}\)
&
\mcell{
A(r)&=1-\frac{2}{r}
+\zeta\frac{\mathcal{E}_r}{r},
\\[1mm]
\mathcal{B}(r)&=1-\frac{2}{r}
+\frac{\zeta\mathcal{E}_3(r-2)(2r-3)}
{r(r-3)^2}
\\
&\quad
-\frac{\zeta\mathcal{E}_r}
{r\beta(r-3)^2}
\Big[
r(r-3)(r-2)
\\
&\qquad
+\beta(r^2-r-3)
\Big].
}
&
\mcell{
A(r)&=1-\frac{2}{r}
+\zeta\frac{\mathcal{E}_r}{r^2},
\\[1mm]
\mathcal{B}(r)&=1-\frac{2}{r}
+\frac{\zeta\mathcal{E}_3(r-2)(2r-3)}
{3r(r-3)^2}
\\
&\quad
-\frac{\zeta\mathcal{E}_r}
{r^2\beta(r-3)^2}
\Big[
r(r-3)(r-2)
\\
&\qquad
+\beta(2r^2-6r+3)
\Big].
}
&
\mcell{
A(r)&=1-\frac{2}{r}
+\zeta\frac{\mathcal{E}_r}{r^3},
\\[1mm]
\mathcal{B}(r)&=1-\frac{2}{r}
+\frac{\zeta\mathcal{E}_3(r-2)(2r-3)}
{9r(r-3)^2}
\\
&\quad
-\frac{\zeta\mathcal{E}_r}
{r^3\beta(r-3)^2}
\Big[
r(r-3)(r-2)
\\
&\qquad
+\beta(3r^2-11r+9)
\Big].
}
\\
\hline

\(\begin{array}{c}X(r)\\ Y(r)\end{array}\)
&
\mcell{
X(r)&=\zeta\left[
\frac{3(\mathcal{E}_3-\mathcal{E}_r)}
{r-3}
+x_1\ln\left(\frac{r}{r-2}\right)
\right],
\\[1mm]
Y(r)&=\zeta\ln\left(\frac{r}{r-2}\right).
}
&
\mcell{
X(r)&=\zeta\left[
\frac{3\left(\frac{\mathcal{E}_3}{3}
-\frac{\mathcal{E}_r}{r}\right)}
{r-3}
+x_1\ln\left(\frac{r}{r-2}\right)
\right],
\\[1mm]
Y(r)&=\zeta\ln\left(\frac{r}{r-2}\right).
}
&
\mcell{
X(r)&=\zeta\left[
\frac{3\left(\frac{\mathcal{E}_3}{9}
-\frac{\mathcal{E}_r}{r^2}\right)}
{r-3}
+x_1\ln\left(\frac{r}{r-2}\right)
\right],
\\[1mm]
Y(r)&=\zeta\ln\left(\frac{r}{r-2}\right).
}
\\
\ 
\\
\hline

\(\begin{array}{c}U(r)\\ V(r)\end{array}\)
&
\mcell{
U(r)&=0,
\\
V(r)&=v_0+v_1\ln\frac{r-2}{r}
-\frac{3}{8\beta^2}
\Bigg[
-\frac{2\mathcal{E}_3^2(r-2)^2\beta^2}{\Delta^3}\Bigg.
\\
&\quad
+\frac{\mathcal{E}_r^2}{\Delta^3}
\Big(2\Delta^2-\Delta(r^2+2r-11)\beta
-2(r-2)^2\beta^2\Big)
\\
&\quad
+\frac{4\mathcal{E}_3\mathcal{E}_r}{\Delta^3}
\Big[-\Delta^2+\Delta(2r-5)\beta
+(r-2)^2\beta^2\Big]
\\
&\quad
+\frac{4\mathcal{E}_3^2(1-2\beta)}{\beta}\Ei_2
+\frac{4\mathcal{E}_3^2(2\beta-1)}{\beta}\Ei_1
\Bigg].
}
&
\mcell{
U(r)&=0,
\\
V(r)&=v_0+v_1\ln\frac{r-2}{r}
-\frac{1}{108\beta^3}
\Bigg[
-2\mathcal{E}_3^2A_\beta\Ei_2
\\
&\quad
+\frac{1}{\Delta^3}
\Big\{-9\mathcal{E}_3^2(r-2)^2\beta^3
+6\mathcal{E}_3\mathcal{E}_r\beta P_2
\\
&\qquad
-\frac{3\mathcal{E}_r^2\beta Q_2}{r^2}
+2\Delta^3
\big(
\mathcal{E}_3^2A_\beta\Ei_1
+\beta B_\beta\Ei_{2r}
\\
&\qquad
-2\mathcal{E}_3\beta^3\Ei_r
\big)
\Big\}
\Bigg].
}
&
\mcell{
U(r)&=0,
\\
V(r)&=v_0+v_1\ln\frac{r-2}{r}
-\frac{1}{972\beta^4}
\Bigg[
\frac{3\beta}{\Delta^3r^4}
\Big\{
-3\mathcal{E}_3^2(r-2)^2r^4\beta^3
\\
&\qquad
+2\mathcal{E}_3\mathcal{E}_r r^3\beta P_3
+\mathcal{E}_r^2Q_3
\Big\}
\\
&\quad
-2\mathcal{E}_3^2\beta C_\beta\Ei_2
+2\mathcal{E}_3^2\beta C_\beta\Ei_1
\\
&\quad
+2D_\beta\Ei_{2r}
-4\mathcal{E}_3\beta^3(2\beta-3)\Ei_r
\Bigg].
}
\\
\hline

\(F[Y]\)
&
\mcell{
F[Y]&=
\frac{\zeta}{\mathcal{R}-3}
\left[
e^{-3/\beta}
-e^{-\mathcal{R}/\beta}
\right].
}
&
\mcell{
F[Y]&=
\frac{\zeta}{\mathcal{R}-3}
\left[
\frac{e^{-3/\beta}}{3}
-\frac{e^{-\mathcal{R}/\beta}}{\mathcal{R}}
\right].
}
&
\mcell{
F[Y]&=
\frac{\zeta}{\mathcal{R}-3}
\left[
\frac{e^{-3/\beta}}{9}
-\frac{e^{-\mathcal{R}/\beta}}{\mathcal{R}^2}
\right].
}
\\
\hline
\end{tabular}
\caption{
Summary of the exponentially corrected nonlocal black hole solutions for
\(n=1,2,3\). We use
\(\mathcal{E}_r=e^{-r/\beta}\), \(\mathcal{E}_3=e^{-3/\beta}\),
\(E=e^{Y/\zeta}\), and
\(\mathcal{R}=2E/(E-1)\). The constants \(v_0\) and \(v_1\) are the
homogeneous integration constants in the reconstructed \(V(r)\). The apparent
\(\Delta^{-3}\) behavior in the displayed \(V(r)\) entries is removable at
\(r=3\), as checked by symbolic series expansion.
}
\label{tab:exp_solutions_compact}
\end{table*}

\section{Analysis of the extended solutions}
\label{sec:graphical_profiles}
We now analyze the behavior of the corrections coming from the logarithmic and exponential terms. In all panels, the reference solution is the Schwarzschild metric. The plotted quantities show the corrections to the temporal component \(A(r)\) and to the physical radial metric coefficient \(B(r)=g_{rr}=1/\mathcal{B}(r)\).

The comparison also illustrates why these correction terms are useful for
phenomenological studies. The logarithmic correction represents a slowly
decaying nonlocal effect, and is therefore more relevant for observations that
probe a wider radial region, such as weak-field lensing. The
exponential correction represents a finite-range effect, and is therefore more
relevant for strong-field observations close to the horizon, while being
suppressed at large radius. The two cases consequently separate two different
physical ways in which nonlocal gravity can modify black-hole observables.

\begin{figure*}[t]
\centering
\begin{minipage}{0.48\textwidth}
\centering
\includegraphics[width=\linewidth]{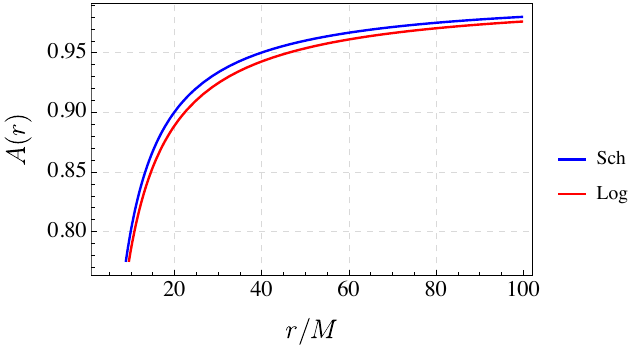}
\par\smallskip
\textbf{(a)} Logarithmic correction to \(A(r)\).
\end{minipage}
\hfill
\begin{minipage}{0.48\textwidth}
\centering
\includegraphics[width=\linewidth]{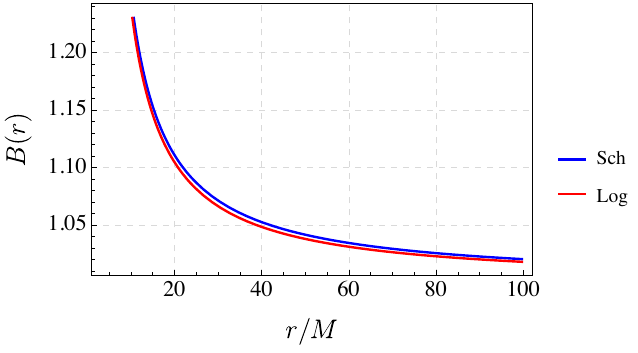}
\par\smallskip
\textbf{(b)} Logarithmic correction to \(B(r)\).
\end{minipage}
\vspace{0.5cm}
\begin{minipage}{0.48\textwidth}
\centering
\includegraphics[width=\linewidth]{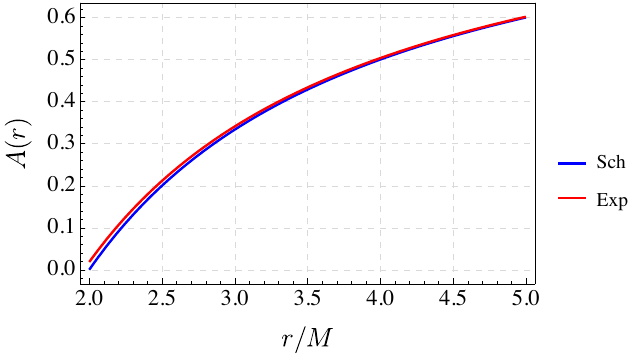}
\par\smallskip
\textbf{(c)} Exponential correction to \(A(r)\).
\end{minipage}
\hfill
\begin{minipage}{0.48\textwidth}
\centering
\includegraphics[width=\linewidth]{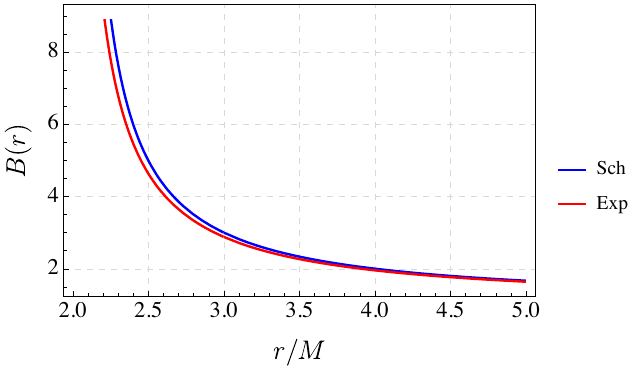}
\par\smallskip
\textbf{(d)} Exponential correction to \(B(r)\).
\end{minipage}
\caption{Metric profiles for the logarithmic and exponential corrections with \(n=1\). Panels (a) and (b) show the logarithmic correction, while panels (c) and (d) show the exponentially suppressed correction. The Schwarzschild result is included as the reference curve in each panel.}
\label{fig:metric_profiles}
\end{figure*}
Panel (a) displays the temporal metric function for the logarithmically dressed
correction,
\begin{equation}
    A_{\log}(r)
    =
    1-\frac{2}{r}
    -\zeta\frac{\ln(r/2)}{r^n}.
\end{equation}
For \(r>2\) and \(\zeta>0\), the logarithmic factor is positive, and therefore
\begin{equation}
    A_{\log}(r)-A_{\Sch}(r)
    =
    -\zeta\frac{\ln(r/2)}{r^n}<0.
\end{equation}
Thus the logarithmic correction lowers \(A(r)\) relative to the Schwarzschild
case throughout the exterior region. Although the correction vanishes
asymptotically, the additional factor \(\ln(r/2)\) delays the approach to the
Schwarzschild limit. Consequently, in the plotted data, \(A(r)\) remains
smaller over a wider radial interval, and the limit \(A(r)\to1\) is reached only
at considerably larger radius.
Panel (b) shows the corresponding radial metric coefficient,
\begin{equation}
    B_{\log}^{\rm phys}(r)=g_{rr}^{\log}(r)=\frac{1}{\mathcal{B}_{\log}(r)}.
\end{equation}
This quantity is sensitive to the correction because the first-order expansion in \(\zeta\) enters the denominator. If
\begin{equation}
    \mathcal{B}_{\log}(r)=A_{\rm Sch}(r)+\zeta\,\delta\mathcal{B}_{\log}(r),
\end{equation}
then, to first order,
\begin{equation}
    \frac{1}{\mathcal{B}_{\log}(r)}
    =
    \frac{1}{A_{\rm Sch}(r)}
    -
    \zeta\frac{\delta\mathcal{B}_{\log}(r)}{A_{\rm Sch}(r)^2}
    +\mathcal{O}(\zeta^2).
\end{equation}
The factor \(A_{\rm Sch}(r)^{-2}\) amplifies the deviation in the strong-field region.
Therefore the logarithmic radial profile approaches the Schwarzschild result
only when \(r\) becomes considerably larger. This explains why the curve for
\(B(r)\) becomes visibly closer to the Schwarzschild curve mainly in the
larger-radius part of the graph.
Panels (c) and (d) display the exponentially suppressed correction,
\begin{equation}
    A_{\exp}(r)
    =
    1-\frac{2}{r}
    +\zeta\frac{e^{-r/\beta}}{r^n}.
\end{equation}
The correction
\begin{equation}
    \Delta A_{\exp}(r)
    =
    \zeta\frac{e^{-r/\beta}}{r^n}
\end{equation}
is largest close to the horizon and is rapidly suppressed at large radius. This
is qualitatively different from the logarithmic correction: the exponential factor
localizes the correction in the near-horizon region. For \(\zeta>0\), the
positive correction raises \(A(r)\) near \(r=2\). As a result, the zero of
\(A(r)\) is shifted inward. Writing
\begin{equation}
    r_H=2+\zeta\,\delta r_H+\mathcal{O}(\zeta^2),
\end{equation}
one obtains
\begin{equation}
    r_H
    =
    2-\frac{\zeta e^{-2/\beta}}{2^{n-1}}
    +\mathcal{O}(\zeta^2).
\end{equation}
Hence the exponential correction makes the horizon radius smaller when
\(\zeta>0\), as shown by the behavior of \(A(r)\) in panel (c).
The same near-horizon localization appears in panel (d), where the plotted
quantity is
\begin{equation}
    B_{\exp}^{\rm phys}(r)=g_{rr}^{\exp}(r)=\frac{1}{\mathcal{B}_{\exp}(r)}.
\end{equation}
Because the physical radial metric component is the reciprocal of \(\mathcal{B}(r)\), even
a small shift in the zero of \(\mathcal{B}(r)\) produces a large effect close to the
horizon. For the exponential correction, the first-order zero of \(\mathcal{B}(r)\) is shifted in
the same inward direction,
\begin{equation}
    r_B
    =
    2-\frac{\zeta e^{-2/\beta}}{2^{n-1}}
    +\mathcal{O}(\zeta^2),
\end{equation}
so the radial pole of \(B(r)=1/\mathcal{B}(r)\) is displaced to a smaller radius for
\(\zeta>0\). Away from the horizon, the exponential factor suppresses the
correction quickly, and the metric rapidly returns to the Schwarzschild form.
Together, the four plots make clear the different physical properties of
the two newly obtained extended solutions. The logarithmic correction produces a slowly decaying deviation,
so both \(A(r)\) and \(B(r)\) require larger radii to become close to their
Schwarzschild limits. The exponential correction, by contrast, is strongly localized
near the horizon; its dominant effect is to shift the horizon inward for
positive \(\zeta\), with both the temporal component \(A(r)\) and the radial
coefficient \(B(r)\) showing the same near-horizon correction pattern.

\section{Discussion and conclusion}\label{sec:conclusion}
In this work, we have extended the static spherically symmetric black-hole solutions of revised D-W nonlocal gravity beyond the single \(r^{-n}\) temporal correction. Since the reduced field equations are linear in the coupling parameter \(\zeta\), as discussed in Sec.~\ref{sec:superposition}, a mode-by-mode superposition of already obtained spherically symmetric solutions can be established. In particular, since the linearity found in the field equations applies to \(\mathcal{B}(r)\), not directly to \(B(r)=g_{rr}\), the obtained solution can produce nonlinear corrections along the radial direction, although it is worth noting that these nonlinear corrections are algebraic consequences of the reciprocal relation and should not be interpreted as independent higher-order solutions.

The physical meaning of this result is that different mechanisms of modified
gravity can be used as generators of nonlocal correction terms in revised D-W
gravity. The \(r^{-n}\) correction is the known correction
of the revised D-W black-hole solution, while the logarithmically
dressed correction is a more direct scale-dependent nonlocal extension that can
originate from quantum effective terms or nonlocal kernels, and the
exponentially suppressed correction can originate from massive scales, spectral
poles, or screening lengths. The superposition structure then allows these
effects to be combined consistently at first order.

Using the same strategy as \cite{DAgostino:2025wgl}, we further constructed two types of asymptotically flat solutions. The logarithmically dressed correction contains the scale \(\beta\) through \(\ell_r=\ln(r/\beta)\), and the exponentially suppressed correction is controlled by the same scale \(\beta\). In both solutions, the apparent singular behavior at \(r=3\) is removable, the normalized scalar function \(W(r)\) is obtained consistently from its formal integral solution, and the auxiliary-sector reconstruction leads to unique solutions of the distortion function \(F[Y(r)]\) and its expression in terms of \(Y\).

The metric analysis shows that the two new solutions have different physical characteristics. For the logarithmically dressed corrections, the correction to \(A(r)\) and the induced correction to the physical radial component \(B(r)\) decay slowly as \(r\) increases, so the return to the Schwarzschild limit is delayed; already in the \(n=1\) case, this can leave a noticeable difference from the Schwarzschild metric over a wider radial interval. For the exponentially suppressed correction, the correction is concentrated near the horizon and, for positive \(\zeta\), shifts the horizon inward at first order. These results show that the revised D-W black-hole equations can admit a broader family of asymptotically flat first-order solutions, with distinguishable strong-field behavior in the temporal and radial metric components.

Therefore, the role of the present construction is not only to add new explicit
metrics, but also to show how physically motivated correction terms from
modified gravity can be incorporated into the D-W nonlocal framework. Paired
with the superposition property, this makes the existing static
spherically symmetric solution more general and provides a useful starting
point for future phenomenological studies of nonlocal black holes.
\begin{acknowledgments}
This work is supported by the National Natural Science Foundation of China (NSFC) with Grant No.~12275087.
\end{acknowledgments}

\bibliographystyle{unsrt}
\bibliography{apssamp}
\end{document}